# Orbital anisotropy underlying the superconducting dome in BaFe$_2$(As$_{1-x}$P$_x$)$_2$ superconductors

T. Sonobe[1], T. Shimojima[1], A. Nakamura[1], M. Nakajima[2], S. Uchida[3] K. Kihou[4], C. H. Lee[4], A. Iyo[4], H. Eisaki[4], K. Ohgushi[5], K. Ishizaka[1]

[1] Quantum-Phase Electronics Center (QPEC) and Department of Applied Physics, The University of Tokyo, Bunkyo, Tokyo 113-8656, Japan.

[2] Department of Physics, Osaka University, Toyonaka, Osaka 560-8531, Japan.

[3] Department of Physics, The University of Tokyo, Bunkyo, Tokyo 113-0033, Japan.

[4] National Institute of Advanced Industrial Science and Technology, Tsukuba 305-8568, Japan.

[5] Department of Physics, Tohoku University, Sendai, Miyagi 980-8578, Japan.

We investigate the in-plane anisotropy of Fe 3$d$ orbitals occurring in a wide temperature and composition range of BaFe$_2$(As$_{1-x}$P$_x$)$_2$ system. By employing the angle-resolved photoemission spectroscopy, the lifting of degeneracy in $d_{xz}$ and $d_{yz}$ orbitals at the Brillouin zone corners can be obtained as a measure of the orbital anisotropy. In the underdoped regime, it starts to evolve on cooling from high temperatures above both antiferromagnetic and orthorhombic transitions. With increasing $x$, it well survives into the superconducting regime, but gradually gets suppressed and finally disappears around the non-superconducting transition ($x \approx 0.7$). The observed spontaneous in-plane orbital anisotropy, possibly coupled with anisotropic lattice and magnetic fluctuations, implies the rotational-symmetry broken electronic state working as the stage for the superconductivity in BaFe$_2$(As$_{1-x}$P$_x$)$_2$.

Iron-based superconductors (*1*) possess a characteristic multi-band structure near the Fermi level ($E_F$), reflecting the five orbitals of Fe 3$d$ electrons. In addition to the complex magneto-structural phase transitions, the lifting of degeneracy in Fe 3$d_{xz}$/3$d_{yz}$ ($xz/yz$) orbitals, the so-called orbital anisotropy or orbital order (*2-4*), was observed across various Fe-pnictide/chalcogenide families. In particular, angle-resolved photoemission spectroscopy (ARPES) has revealed the orbital anisotropy appearing as the imbalance of $xz/yz$ orbitals at the Brillouin zone (BZ) corner, for several classes of materials, such as BaFe$_2$As$_2$ family (*5-7*), NaFeAs (*8-9*), and FeSe (*10-11*). Considering its universality, the orbital anisotropy is now believed to be strongly related to the anomalous normal state such as the electronic nematicity (*8,10-13*), which has been vigorously discussed in terms of magnetic origin (*14,15*). It is also suggested that the orbital fluctuation, which is theoretically known to evolve near the orbital-ordered state (*16,17*), yields a sign-preserved $s_{++}$ wave superconductivity, whereas a sign-changing $s_\pm$ wave superconductivity is likely induced by the spin fluctuation (*18,19*). The nature of the orbital order, in particular to what extent the orbital anisotropy persists with ion-substitution and its relation to the superconducting (SC) dome appearance, can help understand the mechanism of high critical temperature ($T_c$) SC transition in iron-based materials.

The survival of the orbital-anisotropic state, however, has been only investigated for some shallow-doped Ba(Fe$_{1-x}$Co$_x$)$_2$As$_2$ (*6,20*). For precisely tracking the intrinsic orbital anisotropy beyond the SC dome, both materials and methods should be carefully selected. ARPES on strain-free crystals is suitable for avoiding an extrinsic anisotropy increasing the onset $T$ of the orbital order ($T_o$) (*21*). In addition, the isovalent-ion substituted system is most appropriate for thoroughly investigating the degeneracy lifting in the $xz/yz$ orbitals, without the influence of the chemical potential shift induced by the carrier-doping. Thus, we used the strain-free BaFe$_2$(As$_{1-x}$P$_x$)$_2$ (AsP122) crystals with $0.00 \leq x \leq 0.87$ and ARPES for revealing the orbital-anisotropic region in the phase diagram.

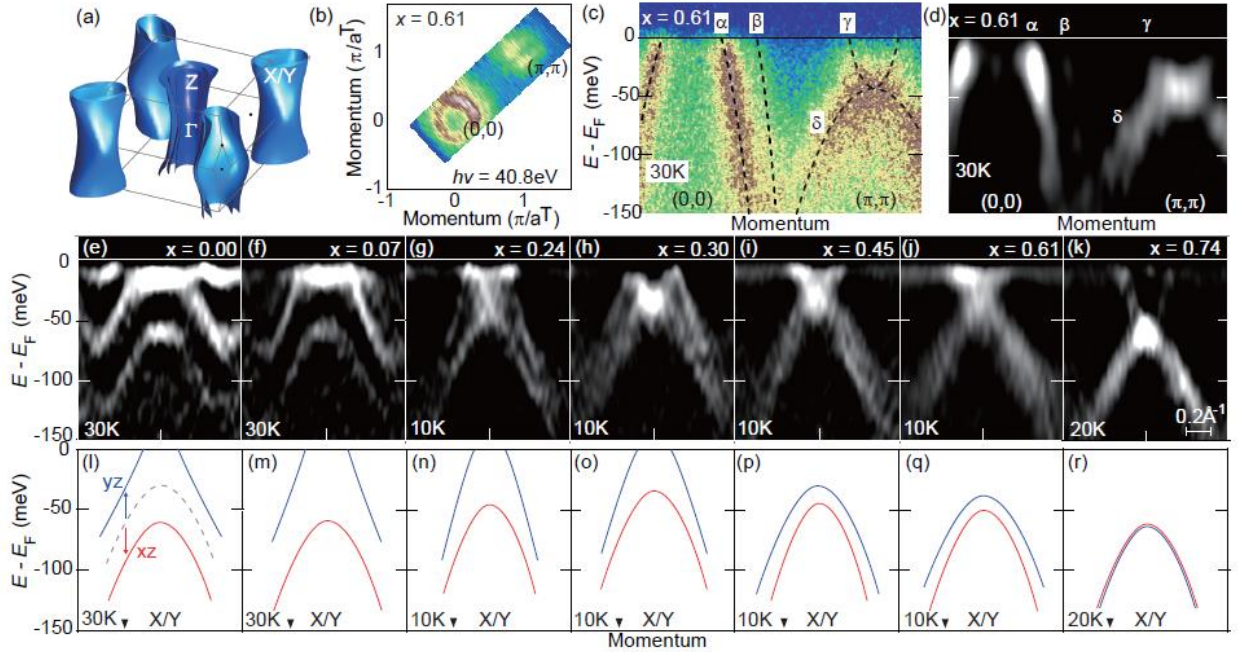

**Figure 1.** (a), First-principles band calculations of FSs for $x = 0.00$. (b), FS mapping at $x = 0.61$ by $h\nu = 40.8$ eV. (c,d), $E$-$k$ image and its second-derivative image for $x = 0.61$ along $(0,0) - (\pi,\pi)$ direction. (e–k), Second-derivative $E$-$k$ images at the BZ corner on twinned crystals of $0.00 \leq x \leq 0.74$. (l–r), Schematic band dispersions obtained from (e) to (k). Blue and red curves correspond to the band dispersions of $yz$ and $xz$ orbitals, respectively [Supplementary Material 1]. Broken curve in (e) shows the degenerate $xz/yz$ band at higher $T$.

Single crystals of AsP122 with $x = 0.00$ ($T_{N,s} = 136$ K), $x = 0.07$ ($T_{N,s} = 114$ K), $x = 0.24$ ($T_{N,s} = 55$ K, $T_c = 16$ K), $x = 0.30$ ($T_c = 30$ K), $x = 0.45$ ($T_c = 22$ K), $x = 0.52$ ($T_c = 15$ K), $x = 0.61$ ($T_c = 9$ K), $x = 0.74$, $x = 0.87$ were grown by a self-flux method as described in Ref. 22 and 23. Resistivity measurements suggest the high quality of the crystals with a residual resistivity ratio of $\leq 30$ (23). ARPES measurements were performed using a VG Scienta R4000WAL electron analyzer and a helium discharge lamp of $h\nu = 40.8$ eV.

Electronic structure of BaFe$_2$As$_2$ is highly two-dimensional as indicated by the calculated Fermi surfaces (FSs) in Fig. 1(a). Figure 1(b) shows the FS image for $x = 0.61$ from ARPES with $h\nu = 40.8$ eV. Energy-momentum ($E$-$k$) image along $(0,0)$ and $(\pi,\pi)$ and its second derivative image are shown in Figs. 1(c) and 1(d). They exhibit two hole bands at the BZ center ($\alpha$ and $\beta$ bands), and an electron band and a hole band at the BZ corner ($\gamma$ and $\delta$ bands). According to the band calculation, the $\delta$ band reaching the $\gamma$ band at 50 meV below $E_F$ is composed of $xz/yz$ orbital characters. Previous ARPES on BaFe$_2$As$_2$ (6) demonstrated that degeneracy lifting in $xz/yz$ orbitals appears in the $\delta$ band in the orthorhombic antiferromagnetic (AF) state.

Here we focus on the $\delta$ band at the BZ corner at low $T$, for the strain-free (twinned) crystals, to estimate the orbital anisotropy $\Phi_o = E_{yz} - E_{xz}$. Figures 1(e)-1(k) show the second-derivative of $E$-$k$ images around the BZ corner for $0.00 \leq x \leq 0.74$, together with the schematics of the band dispersions [Figs. 1(l)-1(r)]. For $x = 0.00$, we observed a pair of hole bands with similar dispersions as indicated by the red and blue curves. Previous ARPES (6) on de-twinned BaFe$_2$As$_2$ crystals confirmed the upward shift of $yz$ hole band and the downward shift of $xz$ hole band, appearing at different BZ corners [$X$ and $Y$ points in Fig. 2(c)]. In the case of twinned crystals, the $\delta$ bands composed of $yz$ and $xz$ orbitals at the $X$ and $Y$ points overlapped in an $E$-$k$ image at the BZ corner. Thus, the observation of the pair of hole bands in the AF state is interpreted as degeneracy lifting in $xz/yz$ orbitals, *i.e.* in-plane orbital anisotropy. The energy difference between these $\delta$ bands gradually decreases with increasing $x$, and disappears around $x = 0.74$ [Figs. 1(l)-1(r)]. These observations indicate that the orbital anisotropy persists toward the over-doped (OD) region where the static orthorhombicity and antiferromagnetism no longer exist.

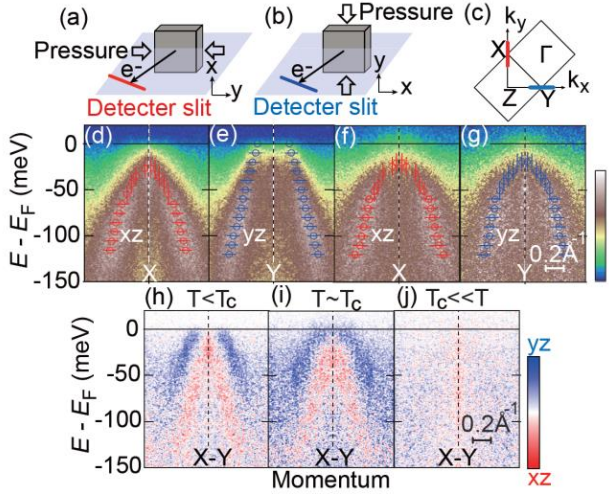
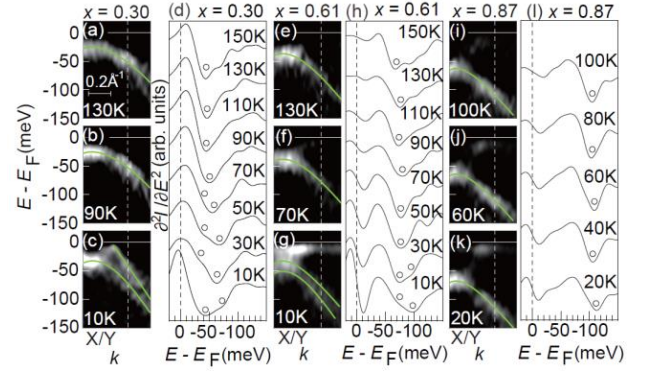

**Figure 2. (a,b),** Geometry 1 and 2 for the ARPES on de-twinned crystals, respectively. **(c),** Momentum cut corresponding to the Geometry 1 (2) as indicated by a red (blue) line. Photons of $h\nu = 40.8$ eV probe the BZ center near the Γ plane and the BZ corner near the Z plane, respectively (35). **(d,e),** ARPES images of de-twinned BaFe$_2$(As$_{0.7}$P$_{0.3}$)$_2$ at 12 K ($T < T_c$) by around $X$ along $k_y$ and $Y$ along $k_x$, respectively. The peak positions from EDCs and MDCs of ARPES images are superimposed by red ($xz$) and blue ($yz$) markers, respectively. **(f,g),** Similar to (d,e) recorded at 150 K ($T \gg T_c$). **(h-j),** Differences between APRES images around $X(\|k_y)$ and $Y(\|k_x)$ at 12 K, 30 K and 150 K, respectively. Red and blue intensities represent $\delta$ bands of $xz$ and $yz$ orbitals, respectively.

**Figure 3. (a-c),** Second-derivative $E$-$k$ images for $x = 0.30$ taken at 130 K, 90 K and 10 K, respectively. Green curves are the guides to the eyes. **(d)** $T$ dependence of the second-derivative EDCs of twinned samples of $x = 0.30$ taken at the momentum indicated by the broken lines in (a-c). Black open circles represent the spectral dips corresponding to the energy position of $\delta$ hole bands. **(e-g),** Similar to (a-c) for $x = 0.61$. **(h),** Similar to (d) for $x = 0.61$. **(i-k),** Similar to (a-c) for $x = 0.87$. **(l),** Similar to (d) for $x = 0.87$.

It is worth mentioning that the signature of the orbital anisotropy is found even in the tetragonal SC state [Figs. 1(h) and 1(i)]. To confirm the degeneracy lifting in $xz/yz$ orbitals in the SC state, we performed ARPES on the de-twinned crystals at $x = 0.30$. As schematically shown in Figs. 2(a) and 2(b), the uniaxial pressure was applied to the single crystals along [1 1 0] direction in the tetragonal notation. This direction is denoted the $y$-axis, corresponding to the shorter $b$-axis in the orthorhombic notation. The $x$-axis corresponds to the $a$-axis in the orthorhombic notation. In ARPES experiments, the crystals were rotated around [0 0 1] *in situ* by 90°, for separately detecting the $\delta$ bands along $k_x$ and $k_y$-directions [Fig. 2(c)].

Figures 2(d) and 2(e) show the band dispersions symmetrized with respect to the $X$ and $Y$ points, obtained for the SC state of $x = 0.30$, respectively. Peak positions of the energy distribution curves (EDCs) and momentum distribution curves (MDCs), plotted in Figs. 2(d) and 2(e), indicate the $\delta$ band attaining higher energy at $Y (\|k_x)$, thus confirming the orbital anisotropy existing in the SC state. First-principles band calculation for BaFe$_2$As$_2$ indicates that the $\delta$ band at $X$ ($\|k_y$) [$Y (\|k_x)$] is composed of $xz$ ($yz$) orbital; thus, these orbitals are marked with red (blue) in Figs. 2(d) and 2(e). The orbital anisotropy, $\Phi_o = E_{yz} - E_{xz}$, is estimated to be ~ 30 meV at the momentum cut 0.3 Å$^{-1}$ away from the $X/Y$ point, which is comparable to the case of twinned crystals [Fig. 1(h)]. These results indicate that the lifting of $xz/yz$ orbital degeneracy in the SC state is not quantitatively affected by the de-twinning procedure. Similar ARPES at 30 K ($T \sim T_c$) demonstrates the comparable magnitude of orbital non-equivalency, suggesting that it is almost insensitive to the superconductivity onset (not shown). At 150 K ($T \gg T_c$), the $\delta$ bands composed of $xz$ and $yz$ orbitals become equivalent [Figs. 2(f) and 2(g)], thus recovering the C$_4$ symmetry of the tetragonal lattice in the high $T$ region. To emphasize the separation between $xz$ and $yz$ orbital bands, we present the images indicating the difference of intensities between the $E$-$k$ image at $X (\|k_y)$ and that at $Y (\|k_x)$, recorded at 12 K [Fig. 2(h)], 30 K [Fig. 2(i)] and 150 K [Fig. 2(j)]. While the orthorhombic structural phase has not been reported from X-ray diffraction (XRD) at any $T$ for $x = 0.30$, the present ARPES may imply the existence of some form of orthorhombicity, *e.g.* the local variation or fluctuation of the orthorhombicity as detected for BaFe$_2$As$_2$ by neutron powder diffraction (24).

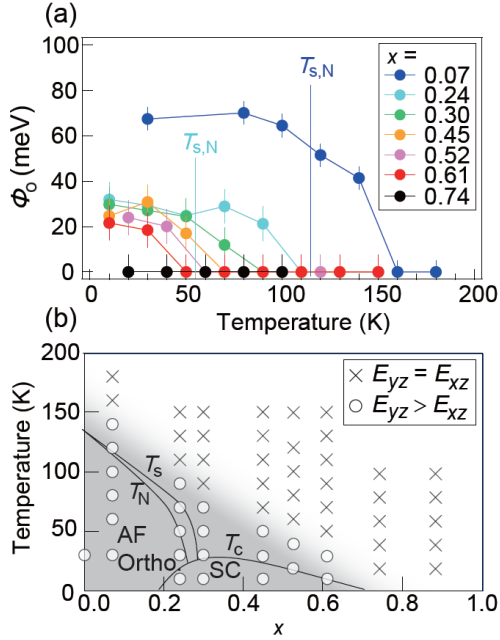

**Figure 4. (a),** Summary of $x$ and $T$ dependence of the orbital anisotropic parameter $\Phi_o = E_{yz} - E_{xz}$ estimated from the second-derivative EDCs. **(b),** The orbital-anisotropic region (shaded area). Circles (crosses) indicate the data points with (without) orbital anisotropy.

To determine the orbital-anisotropic region in the phase diagram, we performed ARPES on AsP122 system for a wide range of $x$ and $T$. We used twinned crystals since the de-twinning procedure may extrinsically increase $T_o$ (21). Figures 3(a)-3(c) show $E$-$k$ images around the $X/Y$ point for $x = 0.30$ obtained at 130 K, 90 K and 10 K, respectively. While a single $\delta$ band is observed at 130 K and 90K, it splits into two at 10 K as indicated by the green curves. To see the $T$ dependence in detail, we show in Fig. 3(d) the second derivative EDCs at the momentum indicated by the dotted white lines in Figs. 3(a)-3(c), plotted for several $T$'s. Spectral dips (open circles) exhibit the energy levels of the $\delta$ band at that momentum. [Supplementary Material 1]. In Fig. 3(d), a single dip gradually splits into two with lowering $T$, indicative of the evolution of the orbital anisotropy. The $T$ at which the two-dip structure appears, corresponding to $T_o$, is determined to be ~90 K for $x = 0.30$. Similarly, $T_o$ can be estimated to be around 50 K for $x = 0.61$ [Figs. 3(e)-3(h)]. At $x = 0.87$, the double-dip feature is no longer observed down to 20 K as shown in Fig. 3(l) [Supplementary Material 2]. These observations are summarized in Fig. 4(a) (25).

As summarized in Figs. 4(b), $T_o$ and $\Phi_o$ of the orbital anisotropy are monotonically suppressed toward the OD region. Considering that both orthorhombic and AF phases show a similar sudden drop immediately below $x \sim 0.30$, the orbital anisotropy seems to be decoupled from these two orders. On the other hand, the orbital-anisotropic region is nearly equivalent to the electronic nematic phase above $T_s$ and $T_N$, detected by the torque magnetometry and XRD (26). Such coincidence might suggest that the loss of rotational symmetry in the lattice and magnetism is induced by the orbital anisotropy and cooperatively evolves from high-$T$ region. Theoretical studies suggested that the imbalance of $xz/yz$ orbitals on the order of 10 meV resolves the magnetic frustration between $(\pi,\pi)$ and $(\pi,-\pi)$, resulting in the enhancement of spin fluctuations (13). Nuclear magnetic resonance studies on AsP122 actually reported the spin fluctuations evolving around $T_o$ in a wide range of $x$, possibly supporting the above scenario (27). It is noteworthy that the orbital-anisotropic region overlaps the pseudo-gap (PG) region observed in ARPES (7) and optical (28) measurements. Theoretical studies predict that spin-nematicity induces the PG in the density of states significantly above $T_s$ (29,30). While the origin of electronic nematicity in AsP122 should be further investigated, these results suggest that the superconductivity in AsP122 emerges at the PG state where the $C_4$ rotational symmetry is broken in lattice, magnetism and orbital degrees of freedoms.

It is also worth noting that the orbital anisotropy seems to disappear simultaneously with the superconductivity at $x \approx 0.7$. In the BaFe$_2$As$_2$ systems, the SC dome disappearance has been discussed in terms of the breakdown of the FS nesting condition due to the carrier doping (31). The origin of the superconductivity disappearance around $x \approx 0.7$ is still under debate since the quasi-nested FSs in the isovalent-ion doped AsP122 remain in a wide range of $x$ (32,33). According to the present ARPES results, the orbital anisotropy and/or related anisotropy in the lattice and magnetism may be strongly related to the expanse of the SC dome in the overdoped AsP122. We note, however, that this may be not the universal behavior in the iron-based superconductors, considering that the phase diagrams of FeCo122 and recently reported Fe(Se,S) systems exhibit $T_o$ ($\approx T_s$) curves intersecting the SC dome, as observed by ARPES studies (6,34). The variety of the orbital-anisotropic region in the normal state will be a key ingredient to systematically understand the diverse phase diagrams and SC gap symmetries in the iron-based superconductors.

In conclusion, we investigated the orbital anisotropy in the AsP122 system by employing ARPES. We found that the orbital anisotropy widely exists in the phase diagram, including the SC regime. It gets monotonically suppressed toward the OD region and eventually disappears almost simultaneously with the extinction of the SC at $x \approx 0.7$. The orbital anisotropy, possibly accompanying the anisotropic lattice and magnetism, may be strongly related to the SC dome in AsP122.

and/or fluctuations in the sample quality have no serious effect on the determination of $T_o$ in the present ARPES.

**Acknowledgments:** We thank Y. Matsuda, T. Shibauchi, S. Kasahara, T. Yoshida, A. Fujimori and H. Kontani for valuable discussions. We acknowledge H. Ikeda for valuable discussions and first-principles band calculations. This research was supported by Toray Science Foundation; Precursory Research for Embryonic Science and Technology (PRESTO), Japan Science and Technology Agency; the Photon Frontier Network Program of the MEXT, Japan; and Research Hub for Advanced Nano Characterization, The University of Tokyo, supported by MEXT, Japan.

Supplementary Materials for

Orbital anisotropy underlying the superconducting dome

in BaFe$_2$(As$_{1-x}$P$_x$)$_2$ superconductors

T. Sonobe[1], T. Shimojima[1], A. Nakamura[1], M. Nakajima[2], S. Uchida[3] K. Kihou[4], C. H. Lee[4], A. Iyo[4],
H. Eisaki[4], K. Ohgushi[5], K. Ishizaka[1]

1 Quantum-Phase Electronics Center (QPEC) and Department of Applied Physics, The University of Tokyo, Bunkyo, Tokyo 113-8656, Japan
2 Department of Physics, Osaka University, Toyonaka, Osaka 560-8531, Japan.
3 Department of Physics, The University of Tokyo, Bunkyo, Tokyo 113-0033, Japan.
4 National Institute of Advanced Industrial Science and Technology, Tsukuba 305-8568, Japan.
5 Department of Physics, Tohoku University, Sendai, Miyagi 980-8578, Japan.

1. **Data analysis of the energy position of *xz* and *yz* orbital bands**

Here we describe the analysis details regarding the determination of the $\delta$ hole band positions in Figs. 1 and 3. Figures S1(a) and S1(b) show the *E-k* image and its second-derivative image near the *X/Y* point at 10 K for *x* = 0.30, respectively. Owing to the presence of twined domains, the $\delta$ bands composed of *yz* and *xz* orbitals were observed in the *E-k* image as indicated by the blue and red curves, respectively. Figure S1(c) exhibits the raw EDC and its second-derivative plot at the momentum indicated by a white broken line in Figs. S1(a) and S1(b). We note that the peak at ~75 meV (red broken line) and the shoulder at ~45 meV (blue broken line) in the EDC correspond to the $\delta$ bands composed of *xz* and *yz* orbitals, respectively. The energy position of each band was determined by the position of the local minima in the second-derivative plot as shown by the broken lines in Fig. S1(c).

Based on the $\delta$ hole band positions (the dip positions in the second-derivative EDCs), we obtained the schematic band dispersions in Figures 1(l)-1(r). Figures S2(a)-S2(i) show second-derivative EDCs near the *X/Y* point in an interval of 0.025 Å$^{-1}$ at 30 K for *x* = 0.00 and 0.07, 20K for *x* = 0.52, 0.74, 0.87 and 10 K for *x* = 0.30, 0.45 and 0.61, respectively. Systematic parallel shift of the double dip feature confirms degeneracy lifting in *xz/yz* orbitals in the $\delta$ bands for $0.0 \leq x \leq 0.61$. In contrast, single $\delta$ band was observed for *x* = 0.74 and 0.87, which confirms an absence of the orbital anisotropy.

2. ***T* dependence of second derivative EDCs near the *X/Y* point for other compounds**

We show the details of the determination of the $T_o$ for *x* = 0.24, 0.45 and 0.74. Figure S3 exhibits the second derivative *E-k* images and EDCs. Similar to the Figure 3, the orbital anisotropy $\Phi$ and $T_o$ for *x* = 0.24, 0.45 and 0.74 were obtained from Figs. S3 and they are summarized in the Figure 4(a) and 4(b). In order to extract a general trend of the orbital anisotropy in AsP122 regardless of the sample preparation method, we used as-grown crystals for *x* = 0.00, 0.07, 0.30 and annealed crystals for *x* = 0.24, 0.45, 0.52, 0.61, 0.74 and 0.87.

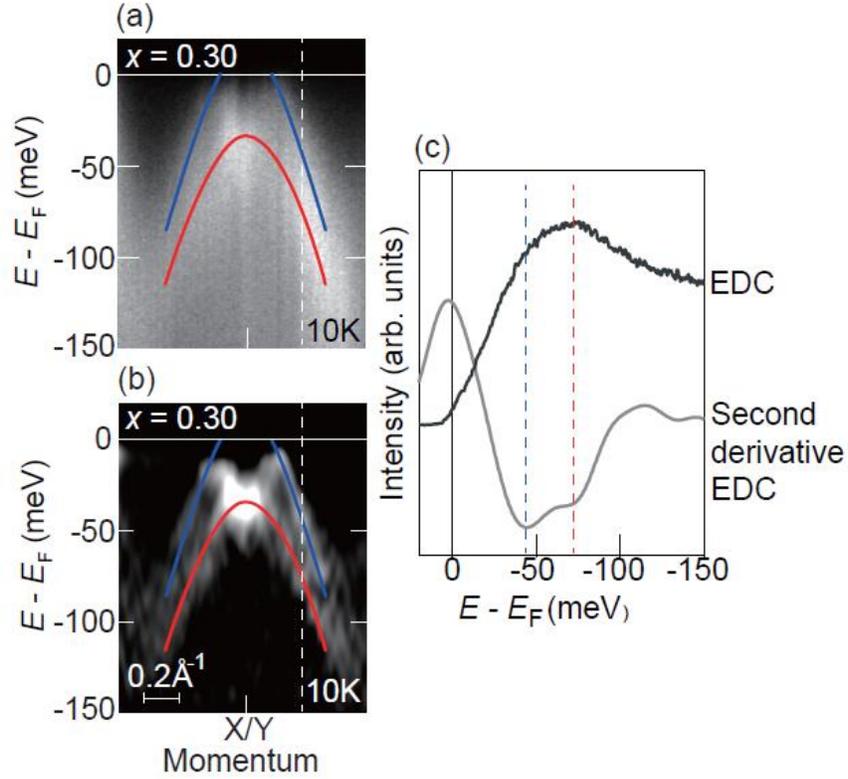

**Figure S1.** (**a,b**) *E-k* image and its second-derivative image of twinned sample for $x = 0.30$ at 10 K along $(0,0) - (\pi,\pi)$ direction, respectively. (**c**), EDC and its second-derivative taken at the momentum indicated by the broken white line in (a) and (b). Blue and red broken lines represent the energy positions of the *yz* and *xz* orbital $\delta$ bands, respectively.

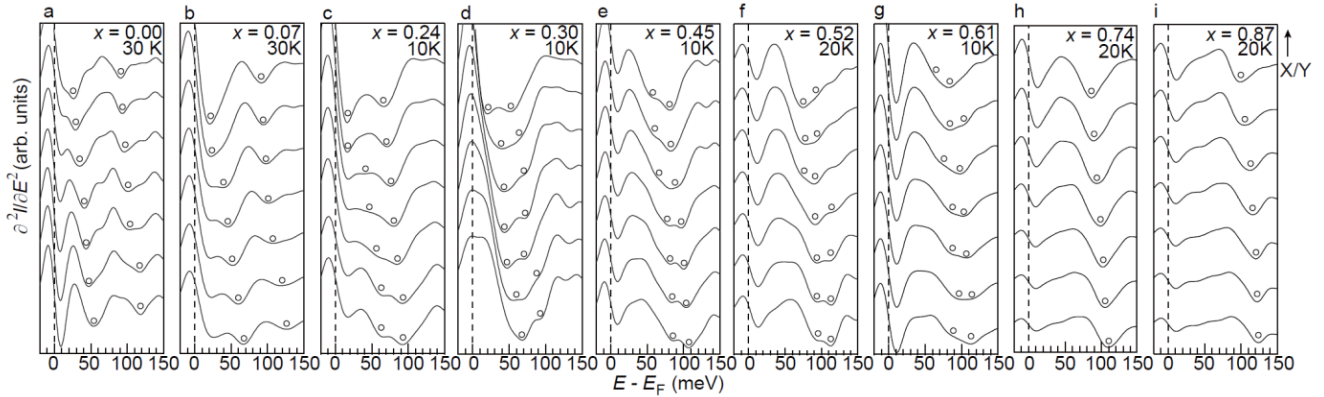

**Figure S2.** (**a-i**), *x* dependence of second-derivative EDCs in an interval of 0.025 Å$^{-1}$ near *X/Y* point at 30 K for $x = 0.00$ and 0.07, 20 K for $x = 0.52$, 0.74 and 0.87, 10 K for $x = 0.24$, 0.30, 0.45 and 0.61, respectively. Black open circles represent dip positions in the spectra corresponding to the energy position of the $\delta$ hole bands.

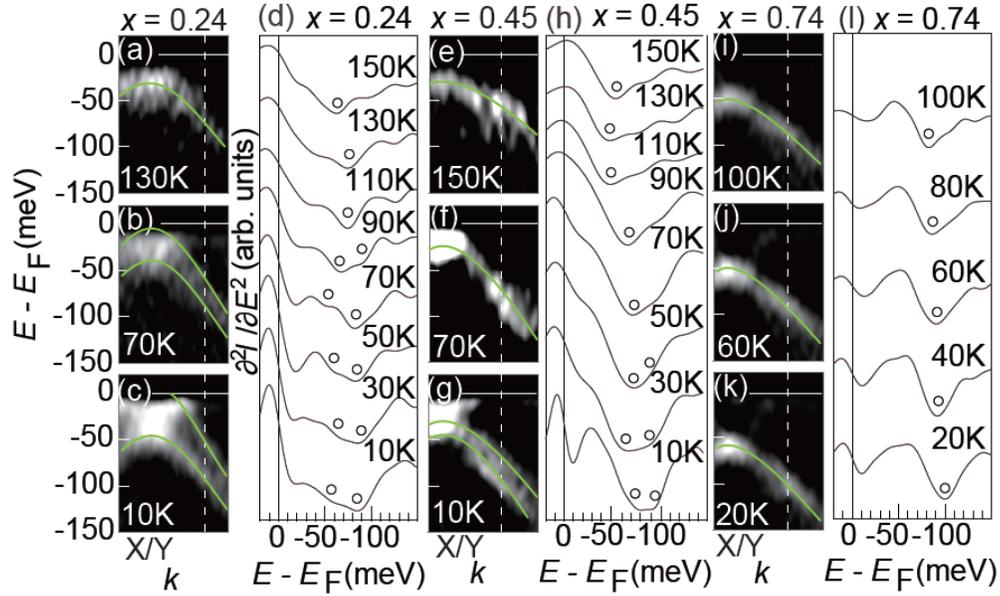

**Figure S3**. (**a-c**), Second-derivative $E$-$k$ images for $x = 0.24$ taken at 130 K, 70 K and 10 K, respectively. Green curves are the guides to the eyes. (**d**) $T$ dependence of the second-derivative EDCs of twinned samples of $x = 0.24$ taken at the momentum indicated by the broken lines in (a-c). Black open circles represent the spectral dips corresponding to the energy position of $\delta$ hole bands. (**e-g**), Similar to (a-c) for $x = 0.45$. (**h**), Similar to (d) for $x = 0.45$. (**i-k**), Similar to (a-c) for $x = 0.74$. (**l**), Similar to (d) for $x = 0.74$.